# Effect of charged impurities on graphene thermoelectric power near the Dirac point

Deqi Wang and Jing Shi

Department of Physics and Astronomy, University of California, Riverside, CA 92521

Abstract

In graphene devices with a varying degree of disorders as characterized by their carrier mobility and minimum conductivity, we have studied the thermoelectric power along with the electrical conductivity over a wide range of temperatures. We have found that the Mott relation fails in the vicinity of the Dirac point in high-mobility graphene. By properly taking account of the high temperature effects, we have obtained good agreement between the Boltzmann transport theory and our experimental data. In low-mobility graphene where the charged impurities induce relatively high residual carrier density, the Mott relation holds at all gate voltages.

Since the first discovery of gated two-dimensional atomic carbon layer device in 2004 [1], tremendous effort has been put into the research of single- and few-layer graphene materials [2-5]. In addition to the electrical conductivity, thermoelectric power or TEP, which is the derivative of the energy-dependent electrical conductivity in the degenerate limit according to the Mott relation [6], provides a sensitive probe to study the transport properties of graphene since the Fermi energy can be conveniently tuned by a gate voltage as reported previously [7-12]. From the temperature dependence of TEP, one can in principle distinguish different scattering mechanisms [10]. Although the Mott relation was used earlier in single-layer [8] and bi-layer [13] graphene systems, in this work, we have prepared graphene devices with a wide range of carrier mobility therefore with a varying degree of disorders, and carefully examined the validity of the Mott relation as we approach the low-density region near the Dirac point.

Single-layer graphene sheets are exfoliated from either Kish graphite or HOPG and selected with optical microscopy followed by electron beam lithography as described in [8]. The inset of Figure 1 is a false colored scanning electron micrograph of a single-layer device for both electrical conductivity $\sigma$ and TEP measurements. For TEP, a temperature gradient, $\nabla T$, is generated by a micro-fabricated heater, resulting in a thermo-voltage response, $V_{th}$. Electrodes 4 & 1 are the current leads, and electrodes 2 & 3 are the voltage leads for measuring both $\sigma$ and $V_{th}$. This four-point (4P) geometry allows us to exclude the contact resistance and to ensure both $\sigma$ and $V_{th}$ to come from the same locations, where the local temperatures are measured by electrodes 2 & 3 via their 4P resistivity as described in [8]. The measurements are carried out in a continuous flow cryostat with a temperature range from 4 to 300 K. The results reported in this works are based measurements on 13 samples, with the carrier mobility $\mu_c$ ranging from 1,500 to 13,000 cm$^2$/Vs. Most graphene samples have both four electrodes plus a heater, but some have two electrodes plus a heater, and some have only four electrodes for electrical measurements.

Fig. 1 shows the measured Seebeck coefficient, $S_M = \frac{E_x}{(\nabla T)_x} = -\frac{\Delta V}{\Delta T}$, of a device with $\mu_c \sim 1{,}500$ cm$^2$/Vs as a function of the gate voltage $V_g$ represented by the open circles. Other three curves are calculated from the measured $\sigma = \sigma(V_g)$ using the Mott relation,

$$S_C^{Mott} = -\frac{\pi^2 k_B^2 T}{3|e|}\frac{\partial \ln\sigma}{\partial \mu} = -\frac{\pi^2 k_B^2 T}{3|e|}\frac{\partial \ln\sigma}{\partial V_G}\frac{\partial V_G}{\partial \mu}, \qquad (1)$$

where $k_B$, $\mu$, and $e$ are the Boltzmann constant, the chemical potential, and the electron charge, respectively. To compute $S_C^{Mott}$, we use $\frac{\partial V_g}{\partial \mu} = \frac{e}{C}\frac{\partial n_{2D}}{\partial \mu}$, where the capacitance per unit area $C$ is 115 $aF/\mu m^2$ for our device geometry, and $\frac{\partial n_{2D}}{\partial \mu}$, the single-particle density-of-states, is determined from the dispersion relation. Three calculated curves, $S_C^{Mott}$, are shown in Fig. 1. First of all, a quadratic dispersion produces the largest discrepancy with $S_M$ (dashed), even using the 4P resistivity which does not include the contact resistance (~ 4,500 Ω). Using a linear dispersion relation, we calculate $S_C^{Mott}$ from $\sigma$ measured with both the two-point (dotted) and 4P (solid) methods and yield better agreement with $S_M$. The best agreement is reached with the 4P resistivity, suggesting that the Mott relation holds for all $V_g$ if the graphene resistivity is properly measured by the 4P method and a linear dispersion relation is used.

We find that the local resistive thermometry reports a larger $\nabla T$, which is probably caused by the high thermal conductivity of graphene [14-15]. This occurs because the local thermometers, i.e. segments of Au/Cr electrodes, are actually much longer than the width of the graphene device (as shown in the inset of Fig. 1); therefore, the temperature rise of the thermometers is primarily determined by the substrate, which consequently overestimates $\nabla T$ of graphene and underestimates the magnitude of $S_M$. We have verified this by comparing the resistance change of the thermometers with that of the graphene probed between electrodes 1&2, 2&3, and 3&4. The discrepancy in the resulting $\nabla T$ evaluated from these two methods can be as large as a factor of two. However, for a fixed temperature, $\nabla T$ should remain constant as $V_g$ is swept; therefore, the measured and calculated TEP should only differ by a $V_g$-independent factor. In Fig. 1, we allow an adjustable parameter to match the calculated TEP curves with $S_M$. The solid curve clearly matches the data best. If the Fermi velocity of $1 \times 10^6$ m/s is used, the $V_g$-independent calibrator factor is found to be about two.

Although similar satisfactory agreement is found in other low-$\mu_c$ samples, high-$\mu_c$ samples exhibit a quite different behavior. Fig. 2a is the TEP data on a much higher $\mu_c$ sample (~ 13,000 cm$^2$/Vs). $S_M$ shows a more diverging trend with a sharp peak and dip near the Dirac point or the charge neutral point (CNP) at all temperatures. Moreover, the diverging $S_M$ can be very well fitted by $\sim \frac{1}{\sqrt{|V_g - V_D|}}$ on both sides except over the central region bounded by the peak and dip. $\Delta V$, the peak-to-dip width in $V_g$, is about 5 V at 200 K, narrower than that in the low-$\mu_c$ sample, i.e. ~10 V in Fig. 1. Fig. 2b shows the similar Mott relation analysis using a linear dispersion and 4P resistivity for four selected temperatures. At 100 K, $S_M$ and $S_C^{Mott}$ agree well over the whole $V_g$ range. At higher temperatures, a deviation starts to develop near CNP and grows progressively in both the magnitude and $V_g$ range. The same qualitative behaviors are observed in other high-$\mu_c$ samples. Due to the aforementioned uncertainty in local temperature measurements, we also allow a $V_g$-independent factor to match the calculated data with $S_M$ at each temperature. We expect the Mott relation to hold at high $V_g$ where the carriers are degenerate; therefore, we force $S_C^{Mott}$ and $S_M$ to match at the highest $V_g$. However, it is impossible to match the sharp features in $S_C^{Mott}$ by varying the adjustable parameter.

The connection between the magnitude of $\mu_c$ and the deviations from the Mott relation is better seen in Fig. 3a. A comparison is made between $S_M$ and $S_C^{Mott}$ in four samples with different $\mu_c$. All measurements were performed at T=200 K. Evidently, the Mott relation holds for the lowest $\mu_c$ sample, but deviates most significantly in the highest $\mu_c$. Below 100 K, the deviation is insignificant even in the highest $\mu_c$ samples (data not shown). More interestingly, this trend is observed in a device whose $\mu_c$ can be set at different values (Fig. 3b). In our earlier study [16], we reported that $\mu_c$ can be widely tuned using molecule-wrapped nanoparticles which modify graphene's charge environment. Using the same method, $\mu_c$ at 295 K is tuned by a factor of two. The contrast between these two cases confirms that the validity of the Mott relation is intimately related to $\mu_c$.

The Mott relation is obtained from the Boltzmann equation which is applicable for single-electron systems. Failure of the Mott relation could indicate importance of the electron-electron interaction in high-$\mu_c$ samples near CNP. However, the fact that it fails only at higher temperatures argues strongly against such a scenario. On the other hand, the Mott relation is only an approximation for degenerate electron systems when $T$ is far below the Fermi temperature $T_F$. In the language of the linear response theory,

$$S = \frac{L^{12}}{L^{11}}, \text{ where } L^{11} = \ell^{(0)}, L^{12} = -\frac{1}{Te}\ell^{(1)}$$

and $$\ell^{(\alpha)} = \int d\varepsilon (-\frac{\partial f}{\partial \varepsilon})(\varepsilon - \mu)^{(\alpha)} \sigma(\varepsilon) \qquad (2)$$

$f(\varepsilon) = \frac{1}{e^{(\varepsilon-\mu)/k_BT}+1}$ is the Fermi-Dirac distribution function. $L^{11}$ and $L^{12}$ are two coefficients in the linear transport equations and $L^{11}$ is simply the electrical conductivity. If $k_BT \ll \mu$, $\left(-\frac{\partial f}{\partial \varepsilon}\right)$ can be legitimately replaced by the delta-function and the leading order in $S$ yields the Mott relation. However, the carrier density near CNP can be so low that $k_BT \ll \mu$ no longer holds; therefore, the Mott relation is violated. This is what precisely occurs in high-$\mu_c$ graphene because the low-density region near CNP renders $k_BT \ll \mu$ invalid. In low-$\mu_c$ graphene, on the other hand, the charged impurities are bountiful, so are the electron and hole puddles in the vicinity of CNP. In the charged impurity model [17-18], the impurity density $n_{imp}$ determines $\mu_c$ by $\mu_c = \left(\frac{20e}{h}\right)\frac{1}{n_{imp}}$. Although the net charge density can be small near CNP, the residual local charge fluctuation, $n^*$, can be significantly large, which implies the absence of a low-density region near CNP.

We determine $n^*$ by $\sigma_{min} = n^* e \mu_c$ [15] and then calculate other relevant parameters for all devices. As shown in table I, both $n_{imp}$ and $n^*$ can vary by an order of magnitude in samples with various mobility values. As a result, the calculated $T_F$ can be as low as 359 K in the highest $\mu_c$ but as high as 1,458 K in the lowest $\mu_c$. The complete $T_F$ vs. $\mu_c$ data are shown in the inset of Fig. 4. Below, we try to assess this effect in terms of a calculated $V_g$ range. We convert $n^*$ to an effective gate voltage $\Delta V_{Cal}^{Imp}$ using $\Delta V_{Cal}^{Imp} = \frac{|e|}{C} n^*$, which is the equivalent gate voltage that produces the corresponding carrier density $n^*$ electrostatically. Then the region from $-n^*/2$ to $n^*/2$ in residual density defines a region near CNP where the transport is governed by electron and hole puddles, the same source for the $\sigma_{min}$ plateau [19]. The calculated width of this region is plotted in Fig. 4. The triangles represent the data from 9 different devices with various $\mu_c$ values, the squares are the data taken from one device (its Seebeck data were shown in Fig. 2) whose variable $\mu_c$ was obtained by manipulating the charge environment using nanoparticles as described in detail in [16]. These two sets of $n^*$ data are calculated from $\sigma$ of different devices taken under different conditions. Surprisingly, the calculated $\Delta V$ from those two sets of data overlap well with each other when they meet in the intermediate $\mu_c$ range. For comparison, the circles are the width of the central region measured from the peak to dip in TEP. Apparently, this width is slightly larger than that determined from $n^*$, which may be attributed to the somewhat arbitrary criterion in defining the region. The former is obtained by reading off the $V_g$ values at the peak and dip in $S_M$ and the latter is essentially defined by the region of the minimum conductivity plateau. Both decrease in the same trend as $\mu_c$ increases, indicating that the TEP behavior near CNP is governed by the residual local charge density.

If $n^*$ is so large that $T_F \gg T$, we expect the Mott relation to hold. This is indeed the case in low-$\mu_c$ devices. If the opposite is true, the Mott relation is violated, which is the case in high-$\mu_c$ samples. At large $V_g$, the electrostatically induced charge density is high, and so is $T_F$. As $V_g$ approaches CNP, the charge density is low in high-$\mu_c$ samples; therefore, the Mott relation fails. In this low-density central region, it is still possible to calculate the Seebeck coefficient from Eq. 2. At finite temperatures, three factors must be considered: full ($-\frac{\partial f}{\partial \varepsilon}$) function, $T$-dependent chemical potential $\mu(T)$, and the energy dependent kernel function, $\sigma(\varepsilon)$. We adopt Eq. 17 in [20] for $\mu(T)$. $\sigma(\varepsilon)$ can explicitly depend on $T$ via electron-phonon interaction and/or dielectric constant due to screening. Although these effects on the kernel function have been addressed theoretically [20], here we replace $\sigma(\varepsilon)$ in $L^{12}$ by measured $V_g$-dependent conductivity, i.e. $\sigma(V_g) = L^{11}$. In the upper left panel of Fig. 2b, we include two additional calculated curves (open triangles and solid) which correspond to the following approximations: (a) replacing $\sigma(\varepsilon)$ by measured a low-$T$ $\sigma(V_g)$, denoted as $S_{Cal}(a)$; (b) replacing $\sigma(\varepsilon)$ by $\sigma(V_g)$ measured finite-$T$, denoted as $S_{Cal}(b)$. Obviously, $S_{Cal}(a)$ leaves out the explicit $T$-dependence of $\sigma(\varepsilon)$, which inevitably underestimates the effects of temperature in $L^{12}$. $S_{Cal}(b)$ uses the measured finite-$T$ conductivity which already includes the effect of the energy spread in $f(\varepsilon)$ along with other temperature effects such as the screening and phonons. Hence, this latter approximation overestimates the temperature effect. Both approximations yield better agreement with the experimental data than the Mott relation calculations. In comparison, the second approximation appears to be slightly better, which indicates that the effects of screening and phonons on $\sigma$ are important at high temperatures. The other panels in Fig. 3a only contain $S_{Cal}(a)$ curves.

In conclusion, we have studied TEP along with the electrical transport and examined the Mott relation in over a dozen graphene samples with a wide range of $\mu_c$ values. In high-$\mu_c$ samples that have low residual carrier density $n^*$, the Mott relation is violated in the vicinity of CNP, which is in contrast to poor $\mu_c$ samples in which the Mott relation is found to always hold over the entire gate voltage range. Finally, the Boltzmann transport theory taking account of the temperature effects can satisfactorily explain the experimentally measured Seebeck coefficient in low-density electron systems near CNP.

We thank Peng Wei, Wenzhong Bao, Vivek Aji, Vincent Ugarte, Chandra Varma, Qian Niu, Le He, and Yadong Yin for their technical assistance and useful discussions. This work is supported in part by DOE DE-FG02-07ER46351 and NSF ECCS-0802214.

Figure Captions:

Figure 1. (Color online) Comparison of experimentally measured Seebeck coefficient $S_M$ (open circles) and three Seebeck curves $S_C^{Mott}$ calculated from measured electrical conductivity using the Mott relation. The solid line is calculated with the 4P resistivity and a linear dispersion relation; the dotted line is with the two-point (2P) resistivity and a linear dispersion relation; and the dashed line is with the 4P resistivity and a quadratic dispersion relation. $\mu_c$ of this device is ~ 1,500 cm$^2$/Vs. The inset shows a false colored scanning electron microscopy image.

Figure 2. (Color online) (a). Seebeck coefficients of a $\mu_c$ device (~ 13000 cm$^2$/Vs) measured from T = 100 to 250 K (correspongding to the solid curves from bottom to top on the left side). The 4P resistivity data are shown in the inset. (b). Comparison of experimentally measured $S_M$ (solid circles) and calculated Seebeck coefficient at four temperatures. The 4P resistivity and a linear dispersion relation are used for all cases. Blue open squares ($S_C^{Mott}$) are the results calculated from the Mott relation. Open triangles ($S_{Cal}(a)$) are calculated using Eq. 2 with σ(V$_g$) measured at T = 100 K. The solid curve ($S_{Cal}(b)$) for T= 230 K is calculated using Eq. 2 but with σ(V$_g$) measured at T= 230 K.

Figure 3. (Color online) (a). Comparison of experimentally measured ($S_M$, solid circles) and calculated ($S_C^{Mott}$, open squares), Seebeck coefficient for four graphene samples with different $\mu_c$ values (from 2,100 to 13,000 cm$^2$/Vs) at T = 200 K. (b). Comparison of the data from one device with two different $\mu_c$ values (1,500 and 3,300 cm$^2$/Vs).

Figure 4. (Color online) Gate voltage range corresponding to the residual charge density range from –n*/2 to n*/2 in devices with varying $\mu_c$ values. Blue triangles are calculated from the data taken at T= 200 K in 9 different devices, whereas green squares are from one device but with a range of $\mu_c$ values at 20 K. Red circles represent the peak-to-dip gate voltage range in measured Seebeck coefficient. Insets

(a) and (b) show the calculated Fermi temperature $T_F$ and residual carrier density $n^*$ vs. $\mu_c$ for all devices, respectively.

Table I. Carrier mobility $\mu_c$, minimum conductivity $\sigma_{min}$, charged impurity density $n_{imp}$, residual carrier density $n^*$, and the Fermi temperature $T_F$ for five representative graphene devices.

References:

-------------------------------------------------------------------------------------------------

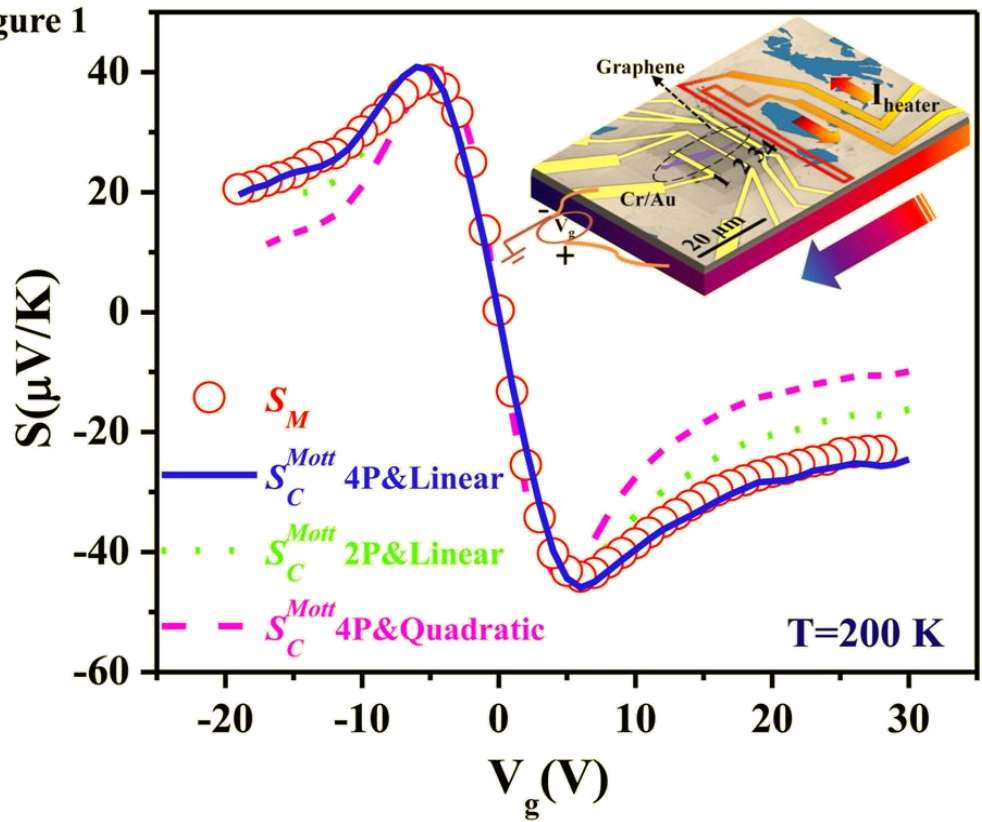

Figure 1

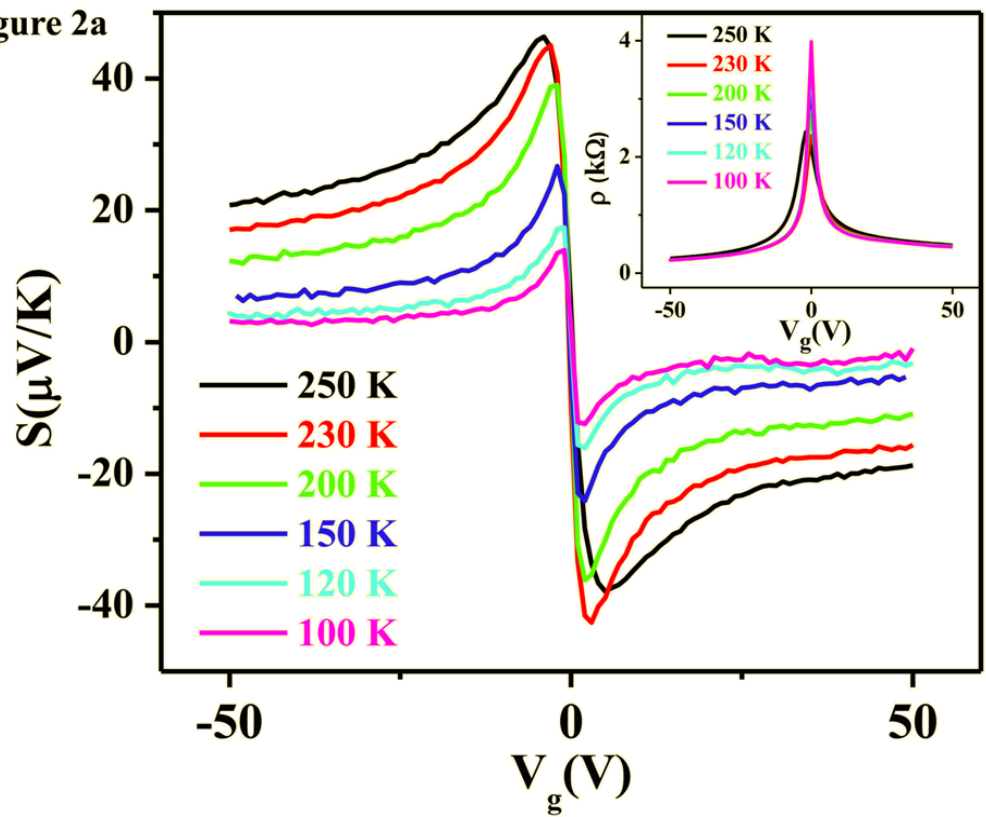

Figure 2a

**Figure 2b**

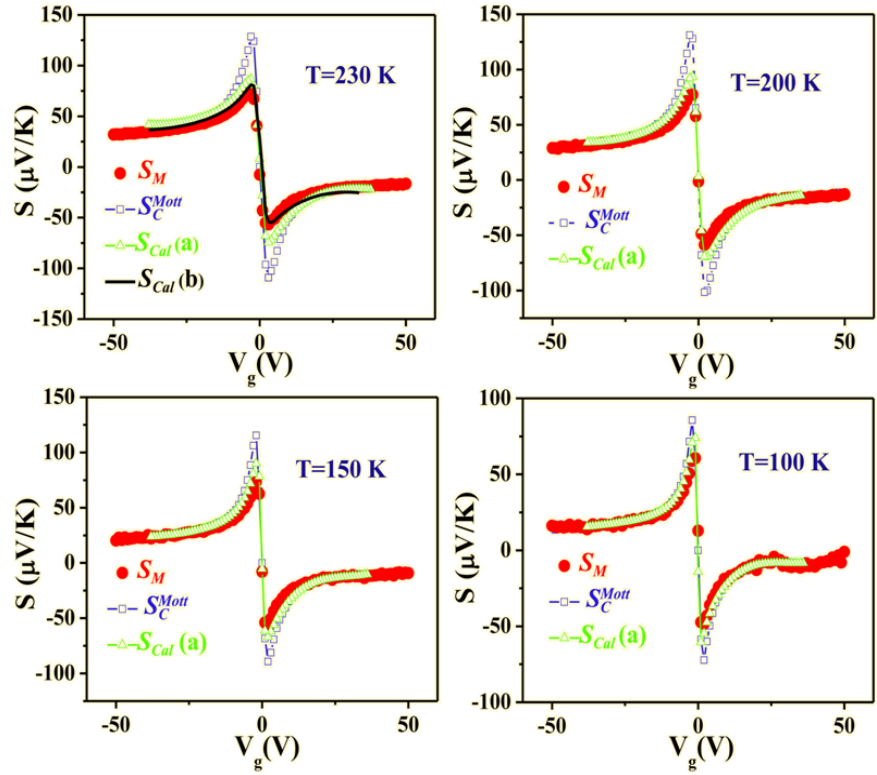

**Figure 3a**

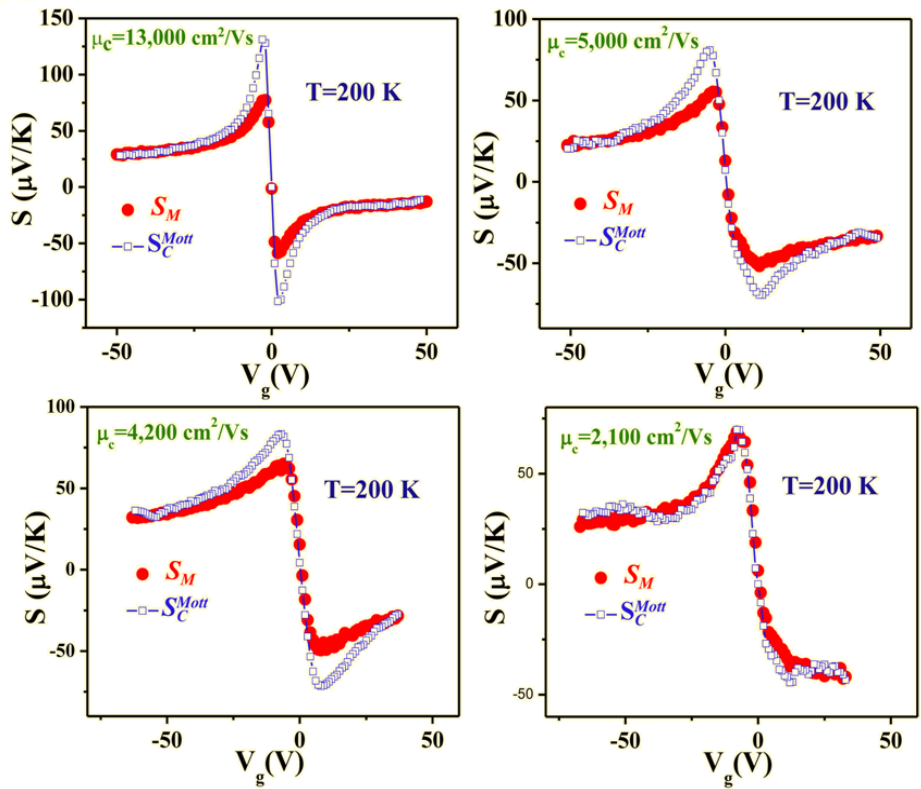

**Figure 3b**

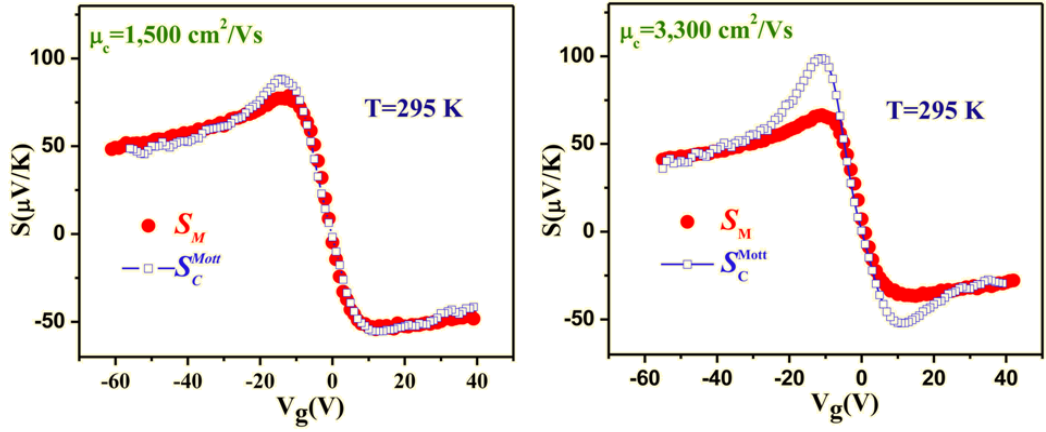

**Figure 4**

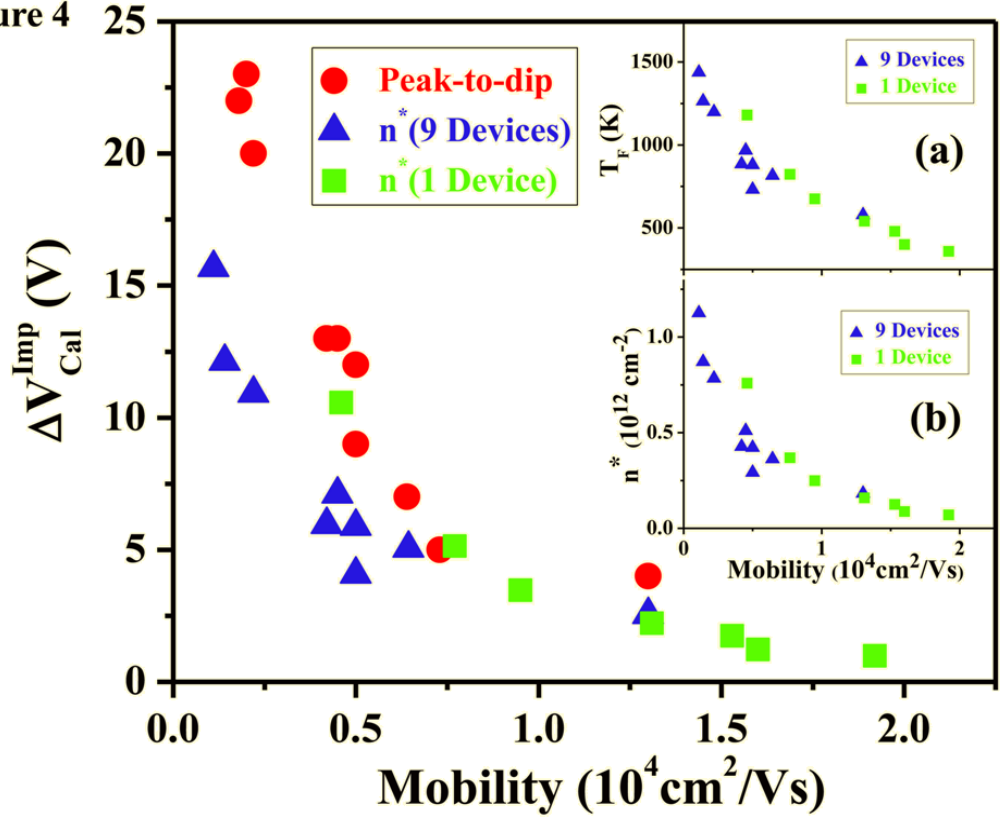

**Table I**

| $\mu_c$ (cm²/Vs) | $\sigma_{min}$ (e²/h) | $n_{imp}$ (cm⁻²) | $n^*$ (cm⁻²) | $T_F$ (K) |
|---|---|---|---|---|
| 13,000 | 9.5 | $3.8 \times 10^{11}$ | $1.8 \times 10^{11}$ | 578 |
| 5,100 | 8.3 | $1 \times 10^{12}$ | $2.9 \times 10^{11}$ | 731 |
| 4,200 | 7.2 | $1.2 \times 10^{12}$ | $4.3 \times 10^{11}$ | 885 |
| 2,100 | 6.9 | $2.3 \times 10^{12}$ | $7.8 \times 10^{11}$ | 1,200 |
| 1,500 | 7.0 | $3.3 \times 10^{12}$ | $1.2 \times 10^{12}$ | 1,458 |